\documentclass[12pt]{amsart}
\usepackage{amsfonts}
\usepackage{amsmath}
\usepackage{amssymb}
\usepackage{color}
\usepackage{graphicx}
\usepackage{xspace}

\newtheorem{prop}{Proposition}[section]
\newtheorem{lem}{Lemma} [section]
\newtheorem{cor}{Corollary}[section]

\newtheorem{thm}{Theorem}[section]
\newtheorem{de}{Definition}[section]

\newcommand{\CC}{\mathbb{C}}

\newcommand{\NN}{\mathbb{N}}

\def \CC{\mathbb{C}}

\begin{document}

\title[A Zassenhaus-type algorithm solves the Bogoliubov recursion] 
      {A Zassenhaus-type algorithm solves \\the Bogoliubov recursion}
      
\author{Kurusch Ebrahimi-Fard}
\address{Laboratoire MIA, 
         Universit\'e de Haute Alsace, 
         4 rue des Fr\`eres Lumi\`ere, 
         68093 Mulhouse, France}
\email{kurusch.ebrahimi-fard@uha.fr}
\urladdr{http://www.th.physik.uni-bonn.de/th/People/fard/}

\author{Fr\'ed\'eric Patras}
\address{Laboratoire J.-A. Dieudonn\'e
         UMR 6621, CNRS,
         Parc Valrose,
         06108 Nice Cedex 02, France}
\email{patras@math.unice.fr}
\urladdr{www-math.unice.fr/~patras}

\date{22 November 2007}

\begin{abstract}

This paper introduces a new Lie-theoretic approach to the computation
of counterterms in perturbative renormalization.  Contrary to the
usual approach, the devised version of the Bogoliubov recursion does
not follow a linear induction on the number of loops.  It is
well-behaved with respect to the Connes--Kreimer approach: that is,
the recursion takes place inside the group of Hopf algebra characters
with values in regularized Feynman amplitudes.  (Paradigmatically, we
use dimensional regularization in the minimal subtraction scheme,
although our procedure is generalizable to other schemes.)  The new 
method is related to Zassenhaus' approach to the
Baker--Campbell--Hausdorff formula for computing products of
exponentials.  The decomposition of counterterms is parametrized by a
family of Lie idempotents known as the Zassenhaus idempotents.  It is
shown, \textit{inter alia}, that the corresponding Feynman rules
generate the same algebra as the graded components of the
Connes--Kreimer $\beta$-function.  This further extends previous work
of ours (together with Jos\'e~M.~Gracia-Bond\'{\i}a) on the connection
between Lie idempotents and renormalization procedures, where we
constructed the Connes--Kreimer $\beta$-function by means of the
classical Dynkin idempotent.
\end{abstract}

\maketitle


\tableofcontents

\thispagestyle{empty}

\section*{Introduction}
\label{sect:introduction}

Renormalization in perturbative quantum field theory (pQFT) is needed
because amplitudes $U(\Gamma)$ associated to Feynman graphs $\Gamma$
are plagued by ultraviolet (UV) divergences.  These divergences in
general demand a regularization prescription, where by introducing
extra parameters, the amplitudes $U(\Gamma)$ become formally finite.
For instance, in dimensional regularization (DR) they become Laurent
series with a polar part encoding the UV divergences.

After regularization, a renormalization scheme has to be chosen.  In
general, the latter is characterized by an operator, here denoted by
$R$, which implements the extraction of the UV divergence of a
regularized Feynman graph amplitude.  In the context of DR the minimal
subtraction scheme operator $R$ picks out the polar part of the
Laurent series expansion of the amplitude $U(\Gamma)$ associated to a
graph $\Gamma$ in the neighbourhood of the physical space-time
dimension of the theory.  In the BPHZ renormalization prescription,
the map $R$ picks out the terms in the Taylor expansion of the Feynman
amplitude about zero momentum, usually up to an order determined by
the overall (superficial) degree of divergence of the corresponding
Feynman graph.  

The naive procedure that would consist in defining the renormalized
amplitude $U_{{\rm{ren}}}(\Gamma)$ of a Feynman graph $\Gamma$ by
subtracting $R(U(\Gamma))$, works only up to $1$-loop order.
Otherwise it leads to non-physical quantities: in particular, locality
could be violated.  The correct definition of the renormalized
amplitude requires the preliminary treatment of the subdivergences
associated to the subgraphs of a given Feynman graph (associated to a
given theory).  The combinatorics of the renormalization process is
encoded in the Bogoliubov preparation map and the Bogoliubov
recursion~\cite{CaswellK1982,collins1984,Lowen1975,Vasilev04,Zimmermann}.

That recursion is the general subject of the present article.  As in
the previous article~\cite{egp2006}, we disclose further properties of
renormalization schemes by methods inspired from the classical theory
of free Lie algebras (FLAs).  Our results build on Kreimer's discovery
of a Hopf algebra structure underlying the process of perturbative
renormalization~\cite{kreimer1998}, and also on the Connes--Kreimer
decomposition of Feynman rules~\cite{ck2000} \`{a} la
Birkhoff--Wiener--Hopf (BWH).  Here we further exploit the
interpretation of Feynman amplitudes as regularized Hopf algebra
characters in terms of a Lie-theoretic version of Bogoliubov's
recursion.  Indeed, the recursion now takes place inside the group of
Hopf algebra characters (instead, as it is the case for the original
Bogoliubov recursion, in the much larger space of linear functions on
the set of forests of the theory).  We show that, as Hopf algebra
characters, the counterterm and the renormalized amplitude split into
components naturally associated to a remarkable series of Lie
idempotents, known as Zassenhaus idempotents, as well as to another
closely related series, baptized accelerated Zassenhaus idempotents.
The first series of idempotents, introduced by Krob, Leclerc and
Thibon~\cite{klt1997} in the setting of noncommutative symmetric
functions, has been more recently studied by Duchamp, Krob and
Vassilieva \cite{dkv2000}.  Their combinatorial and Lie-theoretic
properties can be regarded as reflecting the Zassenhaus formula
familiar from numerical analysis and physics applications.  (In linear
differential equations theory, similar to the Magnus
expansion~\cite{Magnus54,MielPleb70}, which is the continuous analogue
of the Baker--Campbell--Hausdorff series, the Zassenhaus series has a
continuous analogue called the Fer expansion~\cite{EM1,Wilcox67}.)  In
this paper we concentrate on several properties of the Zassenhaus
idempotents relevant for perturbative renormalization in QFT.\\

The article is organized as follows.  We first settle some notation
and recall qualitative features of the Bogoliubov recursion.  We
then introduce directly the new Lie-theoretic version of the
recursion.  The by-product is the decomposition of the regularized
Feynman character, and hence amplitudes, into an infinite product,
whose components are indexed by the number of loops (that is, the
degree in the Hopf algebra of graphs).  In the third section, we
recall the Zassenhaus recursion and its descent algebra
interpretation.  The last section relates this interpretation to the
new recursion.  We show that the exponential components of the
counterterm and renormalized amplitudes (appearing in the new
decomposition) satisfy universal Lie-theoretic properties.  That is,
they are linked to the global counterterm and the renormalized Feynman
rules by means of universal formulas and coefficients pertinent to the
Zassenhaus idempotents ---in much the same way as the Connes--Kreimer
$\beta$-function is underlied by the Dynkin idempotent~\cite{egp2006}.


\section{Rota--Baxter algebras and the Bogoliubov recursion}
\label{sect:RBBogoliubov}

Let us recall the general principles of the Bogoliubov recursion in
the Connes--Kreimer Hopf algebraic approach to renormalization in
pQFT. For details, we refer to the original
articles~\cite{kreimer1998,ck2000}, and their successive refinement
and generalization in terms of Rota--Baxter
algebras~\cite{EGP2,EMP,EM2}.

For a given renormalizable QFT (say, $\phi^4_4$-theory) the
perturbative approach consists of expanding $n$-point Green functions
in terms of Feynman diagrams with $n$~external legs.  The latter are
constructed out of propagators and $4$-valent interaction vertices.
We are not interested in questions concerning the overall convergence
properties of those expansions.  The perturbative expansions for the
$2$-point and $4$-point Green functions are already not well defined
due to the UV divergences appearing in the corresponding Feynman
amplitudes beyond the tree-level.  In the Hopf algebraic approach one
uses the set $\mathcal F$ of one-particle irreducible (1PI) $2$-leg
and $4$-leg Feynman diagrams,\footnote{The notion of $n$-particle
irreducible diagrams corresponds to the maximal number, $n$, of
propagators that can be cut without making the Feynman diagram
disconnected.  We refer the reader to aforementioned references.} to
generate a polynomial algebra $H$ over the complex numbers $\CC$ (the
free commutative algebra over $\mathcal F$).  The algebra $H$ inherits
a graded algebra structure from the decomposition of $\mathcal F$
according to the number of loops of a given diagram.  It receives a
coproduct and a Hopf algebra structure from the process of extracting
1PI UV-divergent subdiagrams out of a given Feynman diagram.  In
summary, Feynman diagrams for any QFT can be organized into a graded
connected commutative Hopf algebra $H$.  We denote its product $m: H
\otimes H \to H$; the coproduct $\Delta: H \to H \otimes H$; the unit
map $u: \CC \to H$; the co-unit $e: H \to \CC$; and the antipode $S: H
\to H$.  They are such that:
\begin{eqnarray*}
	m(H_p \otimes H_q) \subset H_{p+q},\;\;\; \Delta(H_n) \subset
	\sum_{p+q=n} H_p \otimes H_q,\;\; {\rm{ and}}\ \;\;\; S(H_n)
	\subset H_n.
\end{eqnarray*}
A Feynman diagram with $n$ loops belongs to $H_n$, the degree $n$
component of $H$.

The Feynman rules associating to each Feynman graph its corresponding
amplitude, i.e. a multidimensional iterated integral, are seen in the
framework of regularization as a map $\rho$ from $\mathcal F$ to a
commutative algebra $A$ over $\CC$.  Each set of Feynman rules extends
uniquely to a character of $H$, that is, to a multiplicative map from
$H$ to~$A$.  The set of characters $G(A)$ inherits a pro-unipotent
group structure from the graded Hopf algebra structure of $H$ (a
well-known phenomenon in algebraic geometry: as a functor from
commutative algebras to groups, $G$ is the group scheme associated to
the commutative Hopf algebra $H$).  Similarly, the vector space of
linear maps $Lin(H,A)$ inherits an algebra structure from the
coalgebra structure of $H$: the product $\ast$ in $G(A)$ and
$Lin(H,A)$ is called the convolution product and is given by:
$$
\forall f,g \in Lin(H,A):\;\ f \ast g := \pi_A \circ (f \otimes g) \circ
\Delta.
$$
The unit for~$\ast$, written $e$ as well, is the composition of the
unit of~$A$ with the co-unit of~$H$.

When the Feynman amplitudes are the ``bare'' ones, that is, when they
depend on bare coupling constants of the theory, then
$\varphi(\Gamma)=U(\Gamma)$ is ill-defined for a large class of
diagrams $\Gamma \in H$.  In DR together with minimal subtraction one
introduces a regularizing parameter $\epsilon$, a deformation
parameter for the space-time dimension of the theory.  This allows to
define a regularized Feynman rule as a map $\varphi$ from $\mathcal F$
to the field of Laurent series ${\mathcal L}:=\CC
[\epsilon^{-1},\epsilon ]]$.  Now, $\mathcal L$ is a weight one
Rota--Baxter algebra, that is, for any $(x,y) \in {\mathcal L}^2$, we
have:
$$
	R_-(x)R_-(y) = R_-(xR_-(y)) + R_-(R_-(x)y) - R_-(xy)
$$
where $R_-$ is the projection mapping each $x \in \CC
[\epsilon^{-1},\epsilon ]]$ to its strict polar part, i.e. $R_-(x) \in
\epsilon^{-1} \CC [\epsilon^{-1}]$ of $\mathcal L$, orthogonally to
the algebra of formal power series $\CC [[\epsilon]]$.  We set
$R_+:=1-R_-$, so that $R_+$ stands for the projection onto $\CC
[[\epsilon ]]$ orthogonally to $\epsilon^{-1}\CC [\epsilon^{-1}]$.
Hence, the projectors $R_-$ and $R_+$ correspond to a splitting
${\mathcal L}={\mathcal L}_- \oplus {\mathcal L}_+$ into two
subalgebras with $1 \in {\mathcal L}_+$.

\smallskip

The Bogoliubov recursion allows to disentangle the combinatorics of
Feynman graphs.  It is defined in terms of Bogoliubov's preparation
map $\varphi \longrightarrow \overline\varphi$, which sends the group
$G({\mathcal L})$ to the set of $\CC$-linear maps from $H$ to
$\mathcal L$.  It can be defined, together with the counterterm
character $\varphi_-$ and the renormalized character $\varphi_+$ by
the set of equations:
\begin{equation}
	\overline \varphi := \varphi_- \ast (\varphi - e) 
	\;\;\; \ {\rm{ and }}\ \;\;\;
	\varphi_{\pm} = e \pm R_{\pm}(\overline\varphi).
\label{eq:anatema}
\end{equation}
The Rota--Baxter algebra structure on $\mathcal L$ insures that
$\varphi = \varphi_-^{-1} \ast \varphi_+$ as well as that $\varphi_-$
and $\varphi_+$ are characters \footnote{Recall that for $\psi$
a character on $H$ its inverse is given by $\psi^{-1} := \psi \circ S$.}: they are
multiplicative maps respectively from $H$ to $\CC [\epsilon^{-1}]$ (in
fact $\varphi_-$ maps $H^+$ to $\epsilon^{-1} \CC [\epsilon^{-1}]$)
and from $H$ to $\CC [[\epsilon]]$.  In our case scheme, when setting
$\epsilon =0$ the renormalized character induces a well-defined,
scalar-valued character $\varphi_{{\rm{ren}}}$ from $H$ to $\CC$,
which values on Feynman graphs are the renormalized amplitudes of the
theory.

The solution to the system~\eqref{eq:anatema} is unique.  It can be
reached at by induction on the number of loops (hence the terminology
``Bogoliubov recursion'').  We pointed out already that the recursion
takes place in $Lin(H,{\mathcal L})$ and not in $G({\mathcal L})$
---since the preparation map is not a character.  We refer to
\cite{collins1984} for the classical approaches to the Bogoliubov
recursion (including its solution by means of Zimmermann's forest
formula) and to~\cite{EMP} for a recent new approach.

\section{A Lie theoretic decomposition of characters}
\label{sect:LieApproach}

Keep in mind the framework of pQFT. We consider the same
renormalization scheme as before.  In all likelihood, our results can
be extended to more general settings ---e.g. to other Rota--Baxter
target algebras for the group of characters--- but we refrain from
seeking the utmost generality.  Recall that, for an arbitrary
commutative algebra $A$, an $A$-valued infinitesimal character is an
element $\mu$ of the graded Lie algebra $\Xi({A})=
\bigoplus_{n=1}^\infty\Xi_n({A})$ associated to the group $G({A})$,
that is, an element of $Lin(H,A)$ that vanishes on $\CC =H_0$ and on
the square~$(H^+)^2$ of the augmentation ideal of $H$.

\begin{de}
An element $\varphi$ of $\bigoplus_{n\in\NN}Lin(H_n,A)$ is
$n$-connected if and only if it can be written $\varphi = e+\sum_{k
\geq n}\varphi_k$, where $\varphi_k \in Lin(H_k,A)$.  An element
$\rho$ of $\Xi({A})$ is $n$-connected if and only if it can be written
$\rho = \sum_{k \geq n}\rho_k$, where $\rho_k \in Lin(H_k,A)$.
\end{de}

In other terms, $\varphi_i=0$ for $0<i<n$, and equally $\rho_i=0$ for
$0 \leq i < n$.  In the following we often identify implicitly a map
such as $\rho_k$ in $Lin(H_k,A)$ with the corresponding map from $H$
to $A$ (i.e. the one equal to $\rho_k$ on $H_k$ and $0$ on $H_i$, $i
\not= k$).  In particular, we will write $\rho_k \in Lin(H_k,A)
\subset Lin(H,A)$.  From now on in this section $A = \mathcal L$.

\begin{lem}
Let $\varphi$ be an $n$-connected character.  Then, there exist unique
infinitesimal characters $\zeta_{n}^\varphi$ and
$\mu_{n,2n-1}^\varphi$, respectively in $\Xi_{n} ({\mathcal L})$ and
$\bigoplus_{j=n}^{2n-1}\Xi_{j} ({\mathcal L})$ such that:
$$
\exp\bigl(-R_-(\zeta_{n}^\varphi)\bigr) \ast \varphi \ast
\exp\bigl(-R_+(\zeta_{n}^\varphi)\bigr)
$$
is an $n+1$-connected character, respectively
$$
\exp\bigl(-R_-(\mu_{n,2n-1}^\varphi)\bigr) \ast \varphi \ast
\exp\bigl(-R_+(\mu_{n,2n-1}^\varphi)\bigr)
$$
is a $2n$-connected character.
\end{lem}

Indeed, notice that, for $\rho$ any $n$-connected infinitesimal
character, $R_-(\rho),R_+(\rho)$ and $\exp(\rho)$ are $n$-connected
(respectively as infinitesimal characters and as a character).  Notice
also that, by the very definition of the logarithm, for any
$n$-connected characters $\lambda,\beta$ and $\tau$, we have in the
convolution algebra $Lin(H,A)$:
\begin{align*}
\log (\lambda ) &= \sum\limits_{k\in\NN^\ast}\frac{(-1)^{k-1}}{k}
\Bigl(\sum\limits_{m\geq n}\lambda_m\Bigr)^k,
\\
\log (\lambda \ast \tau) &=
\sum\limits_{k\in\NN^\ast}\frac{(-1)^{k-1}}{k}
\Bigl(\sum\limits_{m\geq n}\lambda_m +\sum\limits_{m\geq n}\tau_m +
\sum\limits_{l,m\geq n}\lambda_l \ast\tau_m\Bigr)^k,
\end{align*}
and therefore, since $\lambda_l \ast \tau_m \in Lin(H_{l+m},A) \subset
Lin(H,A)$:
$$
\log (\lambda )_j= \lambda_j \ , \ \log (\lambda \ast \tau)_j =
\lambda_j +\tau_j ,\ j=n,\ldots,2n-1.
$$
It also follows that:
$$
\log (\tau\ast\lambda \ast \beta )_j=\tau_j+\lambda_j+\beta_j ,\
j=n,\ldots,2n-1.
$$
The proof of the lemma follows by setting: $\zeta_{n}^\varphi:=\log
(\varphi )_{n}=\varphi_{n}$ and $\mu_{i}^\varphi:=\log (\varphi
)_i=\varphi_i,\ i=n,\ldots,2n-1$, where we write $\mu_i^\varphi$ for
the degree $i$ component of $\mu_{n,2n-1}^\varphi$.

\begin{prop}
We have the asymptotic formulas in the group of characters:
$$
\varphi = \lim\limits_{n\to\infty}\exp\bigl(\lambda_1^-\bigr) \ast
\cdots \ast \exp\bigl(\lambda_n^-\bigr) \ast
\exp\bigl(\lambda_n^+\bigr) \ast \cdots \ast
\exp\bigl(\lambda_1^+\bigr)
$$
and
$$
\varphi = \lim\limits_{n\to\infty}\exp\bigl(\tau_1^-\bigr) \ast \cdots
\ast \exp\bigl(\tau_n^-\bigr) \ast \exp\bigl(\tau_n^+\bigr) \ast
\cdots \ast \exp\bigl(\tau_1^+\bigr),
$$
with the recursive definitions:
\begin{align*}
\varphi_{\{n-1\}} &:= \exp({-\lambda_{n-1}^-}) \ast \cdots \ast
\exp({-\lambda_1^-}) \ast \varphi \ast \exp({-\lambda_1^+}) \ast
\cdots \ast \exp({-\lambda_{n-1}^+});
\\
\lambda_n^\pm &:= R_\pm \bigl(\zeta_n^{\varphi_{\{n-1\}}}\bigr);
\\
\varphi_{[n-1]} &:= \exp({-\tau_{n-1}^-}) \ast \cdots \ast
\exp({-\tau_1^-}) \ast \varphi \ast \exp({-\tau_1^+}) \ast \cdots \ast
\exp({-\tau_{n-1}^+});
\\
\tau_n^\pm &:= R_\pm \bigl(\mu_{2^{n-1},2^n-1}^{\varphi_{[n-1]}}\bigl).
\end{align*}
\end{prop}

The proof follows from the previous lemma and the pro-unipotent nature
of~$G({\mathcal L})$.

\smallskip

We have constructed, for a set of Feynman rules corresponding to a
perturbatively treated QFT with $\mathcal L$ as algebra of amplitudes,
two decompositions
$$
	\varphi = \varphi_-^{-1}\ast \varphi_+
	       =\varphi_{-(2)}^{-1} \ast \varphi_{+(2)},
$$ 
with 
$$
	\varphi_-^{-1}:=\lim\limits_{n\to\infty}\exp\bigl(\lambda_1^-\bigr)
	                                        \ast \cdots \ast 
	                                       \exp\bigl(\lambda_n^-\bigr)
\;\;\; \ {\rm{ resp. }}\ \;\; 
	\varphi_{-(2)}^{-1}:=\lim\limits_{n\to\infty}\exp\bigl(\tau_1^-\bigr)
	                                              \ast \cdots \ast 
	                                            \exp\bigl(\tau_n^-\bigr)
$$	                                            
and
$$
	\varphi_+:=\lim\limits_{n\to\infty}\exp\bigl(\lambda_n^+\bigr)
	                                    \ast \cdots \ast 
	                                  \exp\bigl(\lambda_1^+\bigr)
\;\;\; \ {\rm{ resp. }}\ \;\; 
	\varphi_{+(2)}:=\lim\limits_{n\to\infty}\exp\bigl(\tau_n^+\bigr)
	                                         \ast \cdots \ast 
	                                       \exp\bigl(\tau_1^+\bigr).
$$

\begin{thm}
We have $\varphi_-=\varphi_{-(2)}$ and $\varphi_+=\varphi_{+(2)}$.
Moreover, the $\varphi =\varphi_-^{-1}\ast \varphi_+$ decomposition
agrees with the decomposition obtained from the Bogoliubov recursion.
\end{thm}

The proof reduces to a simple unicity argument.  Assume that, for
$\phi\in G({\mathcal L})$, $\phi =\phi_-^{-1}\ast \phi_+$, where
$\phi_--e$ and $\phi_+$ are $\epsilon^{-1}\CC [\epsilon^{-1}]$ and
$\CC [[\epsilon]]$-valued, respectively.  Let us write
$\alpha\hat{\ast}\beta:=\alpha\ast\beta -\alpha-\beta$.  We have:
$$
\phi_-\hat{\ast}\phi =\phi_+-\phi_- - \phi
$$
and therefore $\phi_- = e + R_-(\phi_-) = e + R_-(\phi_+-\phi
-\phi_-\hat{\ast}\phi )$ and $\phi_+ = R_+(\phi_+) = R_+(\phi_-+\phi
+\phi_-\hat{\ast}\phi )$. That is:
$$
	\phi_-=e-R_-(\phi + \phi_-\hat{\ast}\phi );
	\;\ \qquad \;\ 
	\phi_+=e+R_+(\phi + \phi_-\hat{\ast}\phi ).
$$
Hence:
$$
 \phi_-=e-R_-(\phi_-\ast (\phi-e) )
	\;\ \qquad \;\ 
	\phi_+=e+R_+(\phi_-\ast (\phi-e)  )
$$
which is (equivalent to) the Bogoliubov recursion formula.  In
particular, both decompositions $\varphi =\varphi_-^{-1}\ast
\varphi_+$ and $\varphi=\varphi_{-(2)}^{-1}\ast \varphi_{+(2)}$ solve
the Bogoliubov recursion.  Since the latter has a unique solution, the
theorem follows.

\section{The Zassenhaus recursions and the descent algebra}
\label{sect:Zassenhaus}

In this section we review the definition and main properties of the
descent algebra, together with its application to the usual Zassenhaus
recursion.  Our presentation of the latter stems from adapting the
noncommutative symmetric functions definitions
in~\cite{klt1997,dkv2000}; the reader is referred
to~\cite{gelfand1995,Patras1994,Patreu2002,Reutenauer1993} for further
details and proofs.  The second series of Zassenhaus idempotents
(termed ``accelerated'') that we introduce below is new, to the best
of our knowledge.

Recall that the tensor algebra $T(X)=\bigoplus_{n\in\NN}T_n(X)$ over a
countable set $x_1, \ldots ,x_n, \ldots$ is, as a vector space, the
linear span of words $y_1 \cdots y_n,\ y_i \in X$ over $X$.  It is
provided with a graded connected cocommutative Hopf algebra structure
defined in terms of concatenation of words as the product $y_1 \cdots
y_n\ \cdot \ z_1 \cdots z_p := y_1 \cdots y_n z_1 \cdots z_p$, and the
unshuffling map as the coproduct $\Delta:T(X) \to T(X)\otimes T(X)$:
$$
\Delta (y_1 \cdots y_n):=\sum\limits_{{I=\{i_1,\ldots,i_k\} \subset
[n]} \atop {k\leq n}} y_{i_1} \cdots y_{i_k} \otimes y_{j_1} \cdots
y_{j_{n-k}},
$$
where $J = \{j_1,\ldots,j_{n-k}\}=[n]-I$ and $[n]= \{1,\ldots,n\}$.  A
word $y_1 \cdots y_n$ is an element of $T_n(X)$, the degree $n$
component of $T(X)$.  There is a natural isomorphism between $T(X)$
and its graded dual $T^\ast (X):= \bigoplus_{n \in \NN}T_n^\ast(X)$
induced by the scalar product on $T(X)$ for which the words form an
orthonormal basis.  This defines another, graded connected
commutative, Hopf algebra structure on $T^\ast(X)$ that can be
described, once again, by means of operations on words: the product is
now the shuffle product, whereas the coproduct is the deconcatenation
of words.  Since $T^\ast(X)$ is a Hopf algebra, the
set~$End(T^\ast(X))$ of linear endomorphisms of~$T^\ast(X)$ is
naturally provided with an associative convolution product~$\ast$.

\begin{de}
The descent algebra $\mathcal D$ is the convolution subalgebra of
$\bigoplus_{n\in\NN} End(T_n^\ast(X)) \subset End(T^\ast(X))$
generated by the graded projections $p_n : T^\ast(X) \longmapsto
T_n^\ast(X)$.
\end{de}

The descent algebra inherits a graduation ${\mathcal
D}=\bigoplus_{n\in\NN}{\mathcal D}_n$ from the graded algebra
structure on $\bigoplus_{n\in\NN} End(T_n^\ast(X) )$ induced by the
convolution product ---since, for $f \in End(T_n^\ast (X)) \subset
End(T^\ast (X))$ and $g \in End(T_m^\ast (X)) \subset End(T^\ast
(X))$, $f \ast g \in End(T_{n+m}^\ast (X))$.  It embeds naturally into
the direct sum of the symmetric group algebras ${\bf S}_\bullet :=
\bigoplus_{n\in \NN}\CC [S_n]$, where $\CC [S_n]$ is viewed as a
subset of $End(T^\ast_n(X))$ by letting $\sigma \in S_n$ act on
$T_n^\ast(X)$ by letter permutation:
$$
	\sigma (y_1 \cdots y_n) := y_{\sigma^{-1}(1)} \cdots y_{\sigma^{-1}(n)}.
$$ 
This embedding property follows, e.g. from the fact that ${\bf
S}_\bullet$ is closed under the convolution product in
$End(T^\ast(X))$; see~\cite{malreu1995}.

\begin{prop}
The degree $n$-component ${\mathcal D}_n$ of the descent algebra is a
subalgebra of $End(T_n^\ast(X) )$ for the composition of maps, denoted
as usual by $\circ$.  It is a free associative algebra over the $p_n$
for the convolution product $\ast$ and carries naturally a Hopf
algebra structure for which the $p_n$ behave as a series of divided
powers; that is, if we write $\delta$ for the coproduct on $\mathcal
D$:
$$
\delta (p_n) = \sum\limits_{i\leq n} p_i \otimes p_{n-i}.
$$
\end{prop} 

The first part of the Proposition is known as the Solomon fundamental
theorem in the case of Coxeter groups of type $A_n$.  It generalizes
to arbitrary (finite) Coxeter groups \cite{solomon1976}.  The second
part follows directly from the computation of convolution products of
the $p_n$ in terms of descent sets of permutations, see
e.g.~\cite[Chap.9]{Reutenauer1993}.

\begin{de}
The Zassenhaus series (or left Zassenhaus series) $(Z_n)_{n\in\NN}$
(respectively the accelerated or left accelerated Zassenhaus series
$(Z_n^{(2)})_{n\in\NN}$) is the series of elements $Z_n\in{\mathcal
D}_n$ (respectively $Z_n^{(2)}\in{\mathcal D}_n$) defined as the
(necessarily unique) solution of:
$$
Id_{T^\ast (X)}=\exp\bigl({Z_1}\bigr) \ast \exp\bigl({Z_2}\bigr) 
	                                      \ast \cdots \ast 
	                                           \exp\bigl({Z_n}\bigr)\ast \cdots,
$$
respectively by:
$$
Id_{T^\ast (X)}=\exp\bigl(Z_1^{(2)}\bigr) \ast
\exp\bigl(Z_2^{(2)}+Z_3^{(2)}\bigr) \ast \cdots \ast
\exp\bigl(Z_{2^n}^{(2)}+ \cdots + Z_{{2^{n+1}}-1}^{(2)}\bigr) \ast
\cdots.
$$
The right Zassenhaus and right accelerated Zassenhaus series are
defined similarly by:
$$
	Id_{T^\ast (X)}= \cdots \ast \exp\bigl(\tilde Z_n\bigr)\ast 
	                 \cdots \ast \exp\bigl(\tilde Z_2\bigr)\ast 
	                             \exp\bigl(\tilde Z_1\bigr),
$$
respectively by:
$$
Id_{T^\ast (X)}= \cdots \ast \exp\bigl(\tilde Z_{2^n}^{(2)}+ \cdots +
\tilde Z_{2^{n+1}-1}^{(2)}\bigr) \ast \cdots \ast \exp\bigl(\tilde
Z_2^{(2)}+Z_3^{(2)}\bigr) \ast \exp\bigl(\tilde Z_1^{(2)}\bigr).
$$
\end{de}

The existence and uniqueness of a solution to these equations follow
by the same arguments as the ones we developed to construct an
exponential solution to the Bogoliubov recursion.  The $Z_n$ and the
$Z_n^{(2)}$ are primitive elements in the Hopf algebra $\mathcal D$
and, equivalently, the $\exp\bigl(Z_n\bigr)$ are group-like elements;
since the logarithm and the exponential map are bijections between
group-like and primitive elements, this follows by induction on $n$
from the equations:
$$
	Z_n = p_n \circ \log \Bigl( \exp\bigl(-Z_{n-1}\bigr) 
	                            \ast \cdots \ast 
	                            \exp\bigl(-Z_1\bigr)\ast Id_{T^\ast (X)}\Bigr),
$$
and
$$
	Z_n^{(2)}=p_n \circ \log \Bigl(\exp\bigl(-Z_{2^m}^{(2)} - \cdots -Z_{2^{m+1}-1}^{(2)}\bigr)
	\ast \cdots \ast \exp\bigl(-Z_1\bigr)\ast Id_{T^\ast (X)}\Bigr),
$$
where $m$ is the highest integer such that $2^{m+1} \leq n$.
Analogous statements hold for the right series.

The link between left and right series can be made explicit as
follows.  Let us recollect first some classical facts.  Let $S$ be the
convolution inverse of $Id_{T^\ast (X)}$ ($S$ is the antipode of
$T^\ast (X)$ in the language of Hopf algebras).  We have:
$$
	Id_{T^\ast (X)}= \cdots \ast \exp\bigl(\tilde Z_n\bigr)
	                        \ast \cdots \ast 
	                        \exp\bigl(\tilde Z_2\bigr)\ast \exp\bigl(\tilde Z_1\bigr), 
$$
so that
$$
	S=Id_{T^\ast (X)}^{-1}
	 =\exp\bigl(-\tilde Z_1\bigr) \ast \exp\bigl(-\tilde Z_2\bigr)
	                              \ast \cdots \ast \exp\bigl(-\tilde Z_n\bigr) \ast \cdots 
$$
However, since $S=Id_{T^\ast (X)}^{-1}$, and since $T^\ast (X)$ is
commutative, $S$ is a multiplicative map: $S(xy)=S(x)S(y)$.  This
follows, e.g. from the observation that $S$, being the convolution
inverse of $Id_{T^*(X)}$, belongs to the group $G(H)$ of $H$-valued
characters of $H$).  Also, we have $S^2 = Id_{T^\ast (X)}$.
Therefore:
\allowdisplaybreaks{
\begin{eqnarray*}
	S\circ S &=& Id_{T^\ast (X)}
	        =S \circ \exp\bigl(-\tilde Z_1\bigr)\ast S\circ\exp\bigl(-\tilde Z_2\bigr)
	               \ast \cdots \ast 
	         S\circ \exp\bigl(-\tilde Z_n\bigr) \ast \cdots\\  
         &=&\exp\bigl(S\circ (-\tilde Z_1)\bigr) \ast \exp\bigl(S\circ (-\tilde Z_2)\bigr)
            \ast \cdots \ast \exp\bigl(S\circ (-\tilde Z_n)\bigr) \ast \cdots 
\end{eqnarray*}}
that is, by uniqueness of the decomposition, $S \circ (-\tilde{Z_i}) = Z_i$.

(For convenience, our conventions for the Zassenhaus series $Z_n$ are
slightly different from the ones in \cite{klt1997,dkv2000}; they write
$\frac{Z_n}{n}$ for our $Z_n$.)

\begin{cor}
The $Z_k$ and the $Z_k^{(2)}$ are dual to Lie quasi-idempotents: that
is, the dual elements in $End(T(X) )$ are, up to multiplication by a
scalar, projections from $T(X)$ onto $Lie(X)=Prim(T(X))$.
\end{cor}

For the Zassenhaus series, the quasi-idempotence property was proved
in~\cite{klt1997,dkv2000}, using a characterization of Lie
quasi-idempotents in the descent algebras \cite{gelfand1995}: the
property follows from the $Z_k$ being primitive elements in $\mathcal
D$.  The same argument applies to the right and right
accelerated Zassenhaus series.

\begin{prop}
The descent algebra is freely generated as an associative algebra by
the series of Zassenhaus elements $Z_n$ (respectively by the series
$Z_n^{(2)}, \tilde Z_n, \tilde{Z}_n^{(2)}$).
\end{prop}

This follows by induction on $n$ from the observation that $(Z_n -
p_n)$ is a noncommutative polynomial in the $Z_i$, $i \leq n$, and
from the fact that the $p_n$ generate freely $\mathcal D$ as an
associative algebra.  The same proof holds for the other series.

\smallskip

Recall that the Dynkin operator (of degree $n$) is the element of
$End(T(X))$ mapping the word $y_1 \cdots y_n$ to the iterated Lie
bracket $[\cdots[y_1,y_2]\ldots,y_n].$ The dual operators $D_n$ on
$T(X)^\ast$, that we still call slightly abusively the Dynkin
operators, belong to the descent algebra and generate freely the
descent algebra as an associative algebra, see
\cite{gelfand1995,egp2006}.

\begin{cor}
The Zassenhaus elements (respectively the Dynkin operators $D_n$) can
be expressed as noncommutative polynomials in terms of the Dynkin
operators (respectively the Zassenhaus elements).
\end{cor}

\section{On counterterms and renormalized characters}
\label{sect:counterermRen}

In the present section we show how the four Zassenhaus series of Lie
quasi-idempotents connect to the renormalization process and the
$\beta$-function in pQFT, in the sense of Connes and
Kreimer~\cite{ck2}.  The series do actually provide universal formulas
to split the counterterm and the renormalized characters into products
of exponentials of infinitesimal characters of increasing degrees.
Since the splittings of $\varphi_-^{-1}$ and $\varphi_+$ follow the
same pattern up to a duality (the two splittings are obtained as
left-to-right respectively right-to-left infinite products), we will
develop only the case of the counterterm; the renormalized character
can be treated similarly, replacing the left Zassenhaus series by the
right ones.

Let us recall first a few facts on Hopf algebras and descent algebras
from~\cite{Patras1994}.  For $H$ an arbitrary graded connected
commutative Hopf algebra, there is an algebra map $\alpha_H$ from
$\mathcal D$ to $End(H)$.  The map is defined by
$$
	\alpha_H (p_{n_1} \ast \cdots \ast p_{n_k}) := p_{n_1}^H \ast
	\cdots \ast p_{n_k}^H,
$$ 
where $p_n^H$ is the projection from $H$ to the $n$-th graded
component $H_n$.  It is an algebra morphism for both the convolution
and the composition product in $\mathcal D$ and $End(H)$
\cite[Thm.II.7]{Patras1994}.  The map also behaves nicely with respect
to the coalgebra structures on $H$ (see \cite{Patreu2002,egp2006}):
namely, for $h,h'$ in $H$ and $f \in \mathcal D$, we have:
$$
	\alpha_H(f)(hh') = \alpha_H(f_{(1)})(h) \alpha_H(f_{(2)})(h')
$$
with the (enhanced) Sweedler notation for the coproduct: $\delta
(f)=f_{(1)}\otimes f_{(2)}$.  From now on, for notational convenience,
we will often omit $\alpha_H$ and identify implicitly an element in $
\mathcal D$ with its image in $End(H)$.  In particular, we will
consider the Zassenhaus series as a series of elements in $End(H)$.

So, let $h$ be an arbitrary element of $H$. From the identity
$$
	Id_{T^\ast (X)} = \exp\bigl(Z_1\bigr) \ast \cdots \ast
	\exp\bigl(Z_n\bigr) \ast \cdots
$$
in $\mathcal D$, and since, by the its very definition,
$\alpha_H(Id_{T(X)})=Id_H$, we obtain:
$$
	Id_{H}=\exp\bigl(Z_1\bigr) \ast \cdots \ast \exp\bigl(Z_n\bigr)
	\ast \cdots
$$
Now, for $\mu \in G(A)$ with $A$ an arbitrary commutative algebra, and
for $f,g \in End(H)$ and $h \in H$, we have:
$$
	\mu(f \ast g)(h) = \mu(f(h^{(1)}) \cdot g(h^{(2)}))
	                 =(\mu \circ f \ast \mu \circ g)(h),
$$
so that, in particular:
$$
	\mu \circ \exp(f) =\exp(\mu\circ f).
$$
It yields:
$$
	\mu=\mu\circ Id_H
	   =\exp\bigl(\mu\circ Z_1\bigr) 
	    \ast \cdots \ast 
	    \exp\bigl(\mu\circ Z_n\bigr) \ast  \cdots 
$$
However, by construction $Z_i \in {\mathcal D}_i$.  Therefore, $\mu
\circ Z_i \in Hom(H_i,A) \subset Hom(H,A)$ or, in other terms, $\mu
\circ Z_i$ is zero on the components $H_k$ of $H$ for $k \not= i$.
Also, for any $i>0$, since $\delta (Z_i)=Z_i\otimes 1+1\otimes Z_i$,
$$
	Z_i(hh')=Z_i(h)\cdot e(h') + e(h)\cdot Z_i(h')
$$
which is equal to $0$ for any $h,h'\in H^+$.  In particular, $\mu\circ
Z_i$ is an infinitesimal character of $H$.

Applying this construction to $\varphi_-^{-1}$, the counterterm
associated to a set of Feynman rules, it ensues:
$$
	\varphi_-^{-1} = \exp\bigl(\varphi_-^{-1}\circ Z_1\bigr)
	                \ast  \cdots \ast 
	                \exp\bigl(\varphi_-^{-1}\circ Z_n\bigr) \ast  \cdots 
$$
with $\varphi_-^{-1}\circ Z_i$ an infinitesimal character in
$Hom(H_i,\epsilon^{-1}\CC[[\epsilon^{-1}]])$.  The uniqueness of such
a decomposition insures that $\varphi_-^{-1} \circ Z_i =\lambda_i^-$.

\begin{prop}
Let $\varphi$ be a Feynman rule character and $\varphi =\varphi_-^{-1}\ast
\varphi^+$ its corresponding decomposition into a counterterm and a
renormalized character.  Then:
$$
	\varphi_-^{-1} =\exp\bigl(\varphi_-^{-1}\circ Z_1\bigr)
	               \ast \cdots \ast 
	               \exp\bigl(\varphi_-^{-1}\circ Z_n\bigr)\ast \cdots 
$$
with $Z_i$ the $i$-th Zassenhaus idempotent.  This decomposition
agrees with the exponential decomposition introduced in the first part
of the article.  Analogous statements hold for the accelerated series;
respectively for the renormalized character and the two right
Zassenhaus series.
\end{prop} 

The $\beta$-function associated to a given Feynman rule character
controls the flow of the renormalized Feynman rule character with
respect to the t'Hooft mass scale, see \cite[Sect.7]{egp2006} for
details.  Algebraically, the $\beta$-function reads:
$$
	\beta = \varphi_- \circ D
$$
or $\beta_n = \varphi_- \circ D_n$ with $D = \sum_n D_n$ the Dynkin
operator.  Since $D_n$ can be written uniquely as a noncommutative
polynomial in the $\tilde Z_i$ (and conversely), there are
coefficients $c_{i_1,\ldots,i_k}$ with:
$$
	D_n =\sum\limits_{i_1+ \cdots +i_k=n, \atop  i_j>0}
	c_{i_1,\ldots,i_k} \tilde Z_{i_1} \ast \cdots \ast \tilde Z_{i_k},
$$
and it follows that
$$
	\beta_n = \sum\limits_{i_1+ \cdots +i_k =n, \atop \ i_j>0}
	 c_{i_1,\ldots,i_k} \varphi_- \circ \tilde Z_{i_1} \ast \cdots \ast \varphi_- \circ \tilde Z_{i_k}.
$$
In other terms, the knowledge of the beta function is
\textit{equivalent} to the knowledge of the (right) exponential
decomposition of $\varphi_-$:
$$
	\varphi_- = \cdots \ast \exp\bigl(\varphi_-\circ \tilde Z_n\bigr)
	            \ast \cdots \ast \exp\bigl(\varphi_-\circ \tilde Z_1\bigr).
$$
However, by the argument preceding the corollary in the previous
section, we also have $-\varphi_- \circ \tilde Z_i =
\varphi_-^{-1}\circ Z_i$, for any $i>0$.  Thus the knowledge of the
$\beta$-function is also algebraically equivalent to that of the left
exponential decomposition of the counterterm Feynman character.

\smallskip

The analysis of the physical meaning of the new processes and
quantities introduced in the present article is postponed to further
work.


\subsection*{Acknowledgments}

We thank the organizers of the 7th International Workshop ``Lie Theory
and Its Applications in Physics'', 18-24 June 2007, Varna, Bulgaria,
for giving one of us (FP) the opportunity to present parts of our
recent research results, at that memorable occasion.  We also like to
thank J.~M.~Gracia-Bond\'{\i}a for useful discussions and a very careful 
reading of the manuscript leading to several improvements. We also 
thank D.~Manchon for useful comments and discussions.





\begin{thebibliography}{99}

\bibitem{CaswellK1982}
    W.~E.~Caswell and A.~D.~Kennedy,
    \textsl{Simple approach to renormalization theory},
    {Phys. Rev. D} 25 (1982) 392--408.
    
\bibitem{collins1984}
    J.~C.~Collins,
    \textsl{Renormalization},
    Cambridge University Press, Cambridge, 1984.

\bibitem{ck2000}
    A.~Connes and D.~Kreimer,
    \textsl{Renormalization in quantum field theory and the Riemann--Hilbert problem I. 
    The Hopf algebra structure of graphs and the main theorem},
    {Commun. Math. Phys.} 210 (2000) 249--273.
    \texttt{arXiv:hep-th/9912092}
    
\bibitem{ck2}
     A.~Connes and D.~Kreimer, 
     \textsl{Renormalization in quantum field theory and the Riemann–Hilbert problem. II. 
     The $\beta$-function, diffeomorphisms and the renormalization group},
     {Commun. Math. Phys.} 216 (2001) 215--241.
     \texttt{arXiv:hep-th/0003188} 

\bibitem{dkv2000}
    G.~Duchamp, D.~Krob and E.~A.~Vassilieva,
    \textsl{Zassenhaus Lie idempotents, $q$-bracketing and a new exponential/logarithm       
    correspondence},
    {J. Alg. Comb.} 12 (2000) 251--277.
    
\bibitem{egp2006}
    K.~Ebrahimi-Fard, J.~Gracia-Bond\'\i a and F.~Patras,
    \textsl{A Lie theoretic approach to renormalization},
    {Commun. Math. Phys.} 276 (2007) 519--549. 
    \texttt{arXiv:hep-th/0609035}
    
\bibitem{EGP2}
	K.~Ebrahimi-Fard, J.~Gracia-Bond\'\i a and F.~Patras,
    	\textsl{Rota--Baxter algebras and new combinatorial identities},
    	{Letters in Mathematical Physics} 81 (2007) 61--75.
    	\texttt{arXiv:math/0701031}
	
\bibitem{EMP}
	K.~Ebrahimi-Fard, D.~Manchon and F.~Patras,
    	\textsl{A noncommutative Bohnenblust--Spitzer identity for Rota--Baxter algebras 
    	solves Bogoliubov's recursion},
    	preprint (2007), \texttt{arXiv.0705.1265}.

\bibitem{EM1}
	K.~Ebrahimi-Fard and D.~Manchon, 
	\textsl{A Magnus- and Fer-type formula in dendriform algebras}, 
	preprint (2007), \texttt{arXiv:0707.0607}.       

\bibitem{EM2}
	K.~Ebrahimi-Fard and D.~Manchon, 
    	\textsl{The combinatorics of Bogoliubov's recursion in renormalization},
    	preprint (2007), \texttt{arXiv:0710.3675}.       

\bibitem{gelfand1995}
    I.~M.~Gelfand, D.~Krob, A.~Lascoux, B.~Leclerc, V.~Retakh and J.-Y.~Thibon,
    \textsl{Noncommutative symmetric functions},
    {Adv. Math.} 112 (1995) 218--348.
    \texttt{arXiv:hep-th/9407124}
     
\bibitem{kreimer1998}
    D.~Kreimer,
    \textsl{On the Hopf algebra structure of perturbative quantum field theories},
    {Adv. Theor. Math. Phys.} 2 (1998) 303--334.
    \texttt{arXiv:q-alg/9707029}

\bibitem{klt1997}
    D.~Krob, B.~Leclerc and J.-Y.~Thibon,
    \textsl{Noncommutative symmetric functions II: Transformations of alphabets},
    {Int. J. of Alg. and Comput.} 7 (1997) 181--264.
    
\bibitem{Lowen1975}
    J.~H.~Lowenstein,
    \textsl{BPHZ renormalization},
    in~{{Renormalization theory}}
    (Proceedings NATO Advanced Study Institute, Erice, 1975),
    NATO Advanced Study Institute Series C: Math. and Phys. Sci.
    Vol.~23, Reidel, Dordrecht, 1976.
    
\bibitem{Magnus54}
    W.~Magnus,
    {\textsl{On the exponential solution of differential equations for a linear operator}},
    {Commun.~Pure Appl.~Math.} 7 (1954) 649--673.    
    
\bibitem{malreu1995}
    C.~Malvenuto and C.~Reutenauer,
     \textsl{Duality between quasi-symmetric functions and the Solomon descent algebra},
     {J. Algebra} 177 (1995) 967--982.    

\bibitem{MielPleb70}
    B.~Mielnik and J.~Pleba\'nski,
    {\textsl{Combinatorial approach to Baker--Campbell--Hausdorff exponents}}
    {Ann. Inst. Henri Poincar\'e A} XII (1970) 215--254.
        
\bibitem{Patras1994}
    F.~Patras,
    \textsl{L'alg\`ebre des descentes d'une big\`ebre gradu\'ee},
    {J. Algebra} 170 (1994) 547--566.
    
\bibitem{Patreu2002}
    F.~Patras and C.~Reutenauer,
    \textsl{On Dynkin and Klyachko idempotents in graded bialgebras},
    {Adv. Appl. Math.} 28 (2002) 560--579.        
    
\bibitem{Reutenauer1993}
    C.~Reutenauer,
    \textsl{Free Lie algebras},
    Oxford University Press, Oxford, 1993.    

\bibitem{solomon1976}
    L.~Solomon,
    \textsl{A Mackey formula in the group ring of a Coxeter group},
    {J. Algebra} 41 (1976) 255--268.
    
\bibitem{Vasilev04}
    A.~N.~Vasilev,
    \textsl{The field theoretic renormalization group in critical behavior theory
    and stochastic dynamics},
    Chapman \& Hall/CRC, Boca Raton, FL, 2004.

\bibitem{Wilcox67}
    R.~M.~Wilcox,
    {\textsl{Exponential operators and parameter differentiation in quantum physics}},
    {J.~Math.~Phys.} 8 (1967) 962--982.

\bibitem{Zimmermann}
    W.~Zimmermann,
    \textsl{Convergence of Bogoliubov's method of renormalization in momentum space},
    {Commun. Math. Phys.} 15 (1969) 208--234.
    
\end{thebibliography}
\end{document}